\documentclass[aip,apl,preprint,superscriptaddress,showpacs,floatfix]{revtex4-1}

\usepackage{graphicx,epsfig,amsmath, amssymb}

\newcommand{\fig}[1]{Fig.~\ref{fig:#1}}

\begin{document}


\title{Contactless measurement of electrical conductance of a thin film of amorphous germanium}

\author{T. S. Mentzel}	
	\email{tamarm@mit.edu}
	\affiliation{Department of Physics, Massachusetts Institute of Technology, Cambridge, Massachusetts 02139}
\author{K. MacLean} 
	\affiliation{Department of Physics, Massachusetts Institute of Technology, Cambridge, Massachusetts 02139}
\author{M. A. Kastner} 
	\affiliation{Department of Physics, Massachusetts Institute of Technology, Cambridge, Massachusetts 02139}


\begin{abstract}
We present a contactless method for measuring charge in a thin film of amorphous germanium (a-Ge) with a nanoscale silicon MOSFET charge sensor.  This method enables the measurement of conductance of the a-Ge film even in the presence of blocking contacts.  At high bias voltage, the resistance of the contacts becomes negligible and a direct measurement of current gives a conductance that agrees with that from the measurement of charge.  This charge-sensing technique is used to measure the temperature- and field-dependence of the conductance, and they both agree with a model of Mott variable-range hopping.   From the model, we obtain a density of states at the Fermi energy of 1.6 $\times$ 10$^{18}$ eV$^{-1}$cm$^{-3}$ and a localization length of 1.06 nm.  This technique enables the measurement of conductance as low as 10$^{-19}$ S.
\end{abstract}


\maketitle

The measurement of electrical conductance enables the discovery of new materials properties and the characterization of materials for applications.  The conventional method for measuring conductance relies on contacting the sample with electrodes, and measuring the current-voltage characteristic.  However, a number of factors can impede this traditional transport measurement.  Various semiconductors form blocking contacts with commonly used electrode materials, making it impossible to measure current.  Highly resistive materials, such as high dielectric constant materials and some films of semiconductor nanocrystals,\cite{Drndic, Mentzel} require extraordinarily large voltages to measure current.  In some cases, a sample can be contaminated or damaged by direct contact with an electrode.  To surmount these obstacles, we have developed a contactless method for measuring charge as an alternative to the conventional current-voltage measurement.

In this Letter, we demonstrate a contactless measurement of electrical conductance in a thin film of a-Ge that forms highly resistive contacts with gold electrodes.  The technique is based on the measurement of charge in the a-Ge by a nanometer scale metal-oxide-silicon field-effect transistor (MOSFET) positioned $\approx$100 nm away from the film.\cite{Petta, Fujisawa, Gustavsson, Elzerman, Martin}  Because the gate of the MOSFET is only $\approx$60 nm wide at its narrowest portion, the conductance of the MOSFET channel is sensitive to electrostatic fluctuations in the vicinty of this constriction.\cite{Ralls}  As such, the MOSFET senses the motion of charge in the a-Ge nearby the constriction.  We are able to measure the conductance of the a-Ge film despite blocking contacts at the interface of the film and gold electrodes.  Moreover, this method enables us to measure resistances greater than 10$^{19}$ $\Omega$ with the application of only 1 V to the film, which would be impossible with traditional methods.  We find that the temperature- and field-dependence of the conductance agree with the model of variable-range hopping with a constant density of states.  From this model, we extract values for the density of states and the localization length, which are consistent with the values reported in the literature\cite{Knotek} and thus validate our measurement method.  This contactless method of measuring electrical conductance can be applied to any thin film for which Ohmic currents are difficult to measure.

\begin{figure}
\setlength{\unitlength}{1cm}
\begin{center}
\includegraphics[width=8.4cm, keepaspectratio=true]{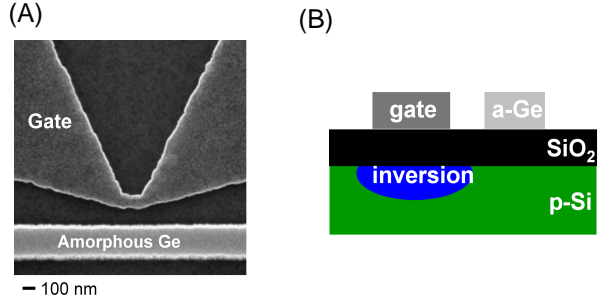}
\end{center}
\caption{(a) An electron micrograph of a film of a-Ge approximately 100 nm away from the narrowest portion of the gate of an \textit{n}-channel Si MOSFET.  Contact is made to the channel of the MOSFET through source and drain electrodes that are not shown here. Gold electrodes are patterned adjacent to the a-Ge film, also not shown, and highly resistive contacts form at the interface as described in the text.  (b) A sketch of the cross sectional view of the device.  The channel, or inversion region, of the MOSFET is electrostatically coupled to the a-Ge film.}
\label{fig:device}
\end{figure}

We fabricate an \textit{n}-channel MOSFET according to standard CMOS methods, as previously described.\cite{MacLean1, MacLean2} We begin with a \textit{p}-type silicon substrate that is doped with boron at a level of 4 $\times$ 10$^{15}$ cm$^{-3}$.  We implant phosphorous ions in two regions of the silicon substrate with dimensions 100 $\times$ 150 $\mu$m$^2$, and separated by 100 $\mu$m, to serve as the source and drain electrodes. A layer of silicon dioxide 100 nm thick is grown via dry, thermal oxidation.  We pattern an \textit{n}$^+$ polysilicon gate on top of the oxide, and use electron beam lithography to define the narrowest portion of the gate, which is $\approx$60 nm wide.  Electrom-beam lithography is used to define an area in polymethyl-methacrylate (PMMA) 2 $\mu$m long and 200 nm wide that is approximately 100 nm away from the narrowest portion of the gate and overlaps with a pair of gold electrodes that are separated by 1 $\mu$m.  A 50 nm thick film of a-Ge is deposited in this region by electron beam evaporation, and the resulting device structure is displayed in \fig{device}.  The device was held in a storage vessel for several months before measurement during which time we expect an oxide to have formed at the interface between the a-Ge and the gold electrodes.

\begin{figure}
\setlength{\unitlength}{1cm}
\begin{center}
\includegraphics[width=7.0cm, keepaspectratio=true]{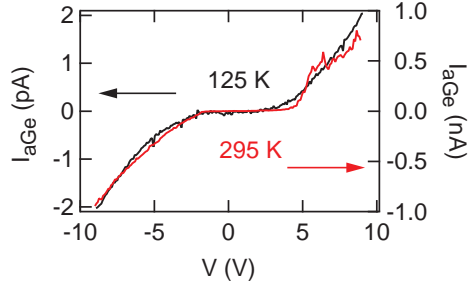}
\end{center}
\caption{Current versus voltage applied to the film of a-Ge at 125 K (black) and 295 K (red).  The non-Ohmic nature of the IV characteristic is indicative of blocking contacts.  The current becomes measurable for applied voltages greater than $\approx$5 V.}
\label{fig:IV}
\end{figure}

The current as a function of voltage for the a-Ge film is displayed in \fig{IV}.  We find that it is non-Ohmic, and in paticular that current is immeasurable for bias voltages less than $\approx$5 V.  This indicates that at low bias the contacts impede current flow, which can be understood as follows.  The aforementioned oxide that is expected to form between the film and the electrodes creates a tunnel junction.  Previous examples of Al-Al$_2$O$_3$-aGe tunnel junctions exhibit a zero-bias anomaly in the tunneling conductance, namely a minimum in the conductance about zero bias.\cite{Fritzsche}  This decrease in the tunneling conductance at zero bias is caused by a Coulomb gap in the density of states at the Fermi energy in the a-Ge film.\cite{CoulombGap}  With a sufficiently high bias voltage, charges tunnel at energies far enough from the Fermi energy that the density of states grows larger, and the tunneling conductance increases.

\begin{figure}
\setlength{\unitlength}{1cm}
\begin{center}
\includegraphics[width=8.4cm, keepaspectratio=true]{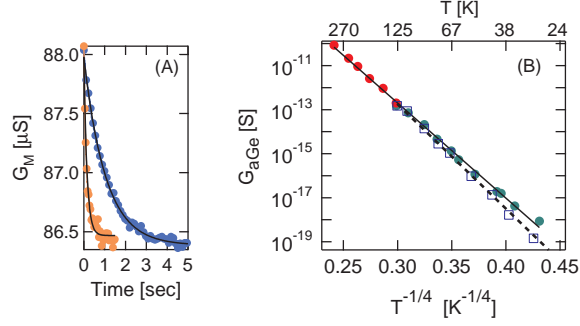}
\end{center}
\caption{(a) The conductance through the Si MOSFET, G$_M$, versus time immediately after changing the voltage bias across the a-Ge film from -0.5 to -1 V at a temperature of 44 K (blue) and 54 K (orange).  The solid lines are fits to an exponential as described in the text. (b) Conductance of the a-Ge film versus temperature derived from measuring charge with a bias voltage of -1 V (blue squares) and -5.5 V (green circles) across the a-Ge film.  At temperatures $\geq$125 K, the conductance is derived from measuring current at a bias of -5.5 V (red cirlces).  Current is immeasurable with a bias of -1 V even at temperatures as high as 295 K.  The lines are fits to Mott variable-range hopping at -1 V (dashed line) and at -5.5 V (solid line) as described in the text.}
\label{fig:TransBlock}
\end{figure}

The contactless technique that we have developed to measure the low-bias conductance in the a-Ge film entails stepping the bias voltage on the a-Ge film so that charge flows in the film, and simultaneously measuring the conductance of the MOSFET as a function of time.  Throughout the measurement, we maintain a positive voltage on the gate of the MOSFET to ensure that the channel is in inversion.  To measure the conductance of the MOSFET, we apply a source-drain voltage of 5 mV and measure the current in the channel.  Initially, both of the electrodes adjacent to the a-Ge are maintained at the same voltage, and then we step the voltage on one of the electrodes to a negative value relative to the other electrode. Even though the oxide prevents measurable current flow between the electrode and the film, capacitive coupling between them causes positive charge in the a-Ge film to flow towards the negatively biased electrode.  The \textit{n}-channel MOSFET senses the decrease in positive charge in the vicinity of the narrow constriction, and the conductance of the MOSFET decreases.  This effect is demonstrated in \fig{TransBlock}(a).  At time t = 0, we step the voltage bias of the a-Ge from -0.5 to -1 V, and the conductance of the MOSFET decreases with time.  We refer to the response of the MOSFET to the charge flow in the a-Ge as a charge transient.

To understand the charge transient, we follow the analysis of MacLean \textit{et. al.}\cite{MacLean1}  When a resistive film with a distributed capacitance is subject to an electric field in a direction parallel to the surface, charge diffuses through the film with a diffusion constant of $D=1/R_{sq}C$ where $R_{sq}$ is the resistance per square and $C$ is the capacitance per unit area of the film.  By solving the diffusion equation for this system, it is found that the charge per unit area at any point in the film varies with time according to $\sigma(t) \approx \sigma_0 + A exp(-\Gamma t)$ where $\Gamma = \pi^2D/L^2$, $L$ is the length of the film, and $D$ is the diffusion constant.  The constants $\sigma_0$ and $A$ depend on the voltage applied to the film and the capacitance of the film.  As long as the voltage step on the a-Ge is sufficiently small, the conductance of the MOSFET varies linearly with the charge in the a-Ge film, so $G_M(t) \approx G_0 + C exp(-\Gamma t)$.  We fit the charge transients in \fig{TransBlock}(a) to this form, and extract $\Gamma$, the rate of charge flow in the film, and hence $R_{sq}$, the resistance per square of the film.  The conductance of the a-Ge film is given by $G_{aGe} = w/R_{sq}L$ where $w$ and $L$ are the width and length of the film respectively.  As shown in \fig{TransBlock}(a), when the temperature of the film is increased from 44 to 54 K, the rate increases, indicating an increase in the conductance of the film.  

With this method, we measure the conductance of the a-Ge film as a function of temperature in a regime where the contacts limit the dc current (\fig{TransBlock}(b), open blue squares).  At each temperature in the range of 30-125 K, we step the voltage on the a-Ge film from -0.5 to -1 V, and derive the conductance from the charge transient of the MOSFET.  We measure conductances smaller than 10$^{-19}$ S with the application of only -1 V.  As expected, the direct current is immeasurable ($<$10 fA) at a bias voltage of -1 V for temperatures as high as 295 K.   

We repeat this set of conductance measurements as a function of temperature in a high-bias regime.   We step the voltage on the a-Ge film from -4.5 to -5.5 V, and derive the conductance from the charge transient of the MOSFET at each temperature in the range of 18-125 K (\fig{TransBlock}(b), green circles).  At temperatures $\geq$125 K, we obtain the conductance of the a-Ge using conventional measurements of current at an applied bias of -5.5 V (red circles).  We fit the conductance over the entire temperature range to a model of variable-range hopping with a constant density of states $G = G_0 exp(-T^*/T)^{1/4}$ as has been found previously.\cite{Knotek}  The fit yields a value for $T^*$ of 9.8 $\times$ 10$^7$ K.  As seen in \fig{TransBlock}(b), we obtain the conductance at 125 K from measurements of both charge and current, and find that the two agree within a factor of 1.3, thereby validating our contactless method for measuring the conductance using the charge transient.  

Knowing that the tranport is described by variable-range hopping, we fit the conductance in the low-bias regime to this model (\fig{TransBlock}(b), dashed line).  When we extrapolate the fit to temperatures greater than 125 K, we find the conductance should be $\geq$10$^{13}$ S. With a 1 V bias, current should be easily measurable ($>$100 fA) and the fact that it is not is consistent with our conclusion that the contacts limit the dc current.

\begin{figure}
\setlength{\unitlength}{1cm}
\begin{center}
\includegraphics[width=7.0cm, keepaspectratio=true]{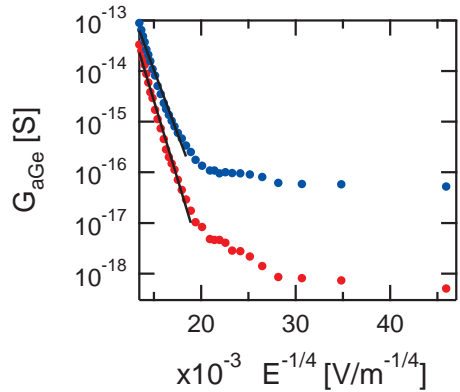}
\end{center}
\caption{Conductance of the a-Ge film versus applied field at a temperature of 33 K (red) and 58 K (blue).  The black lines are fits to a model of variable-range hopping in a strong electric field, as described in the text.}
\label{fig:field}
\end{figure}

The field dependence of the conductance at temperatures of 33 and 58 K is shown in \fig{field}.  We apply an electric field across the a-Ge, and then step the voltage on one electrode by -0.5 V and obtain the conductance from the charge transient.   The black lines in \fig{field} are fits to a model of variable-range hopping in a strong electric field $G = G_0 exp(-E^*/E)^{1/4}$.  We extract a value for $E^*$ of 7.7 $\times$ 10$^{12}$ V/m.  

Given that $T^*=16/k_B D(E_F)a$ and $E^*\approx 16/e D(E_F)a^4$ where $D(E_F)$ is the density of states at the Fermi energy and $a$ is the localization length, we take the values of $E^*$ and $T^*$ from the fits, and find $D(E_F)=1.6\times10^{18}$ eV$^{-1}$ cm$^{-3}$ and $a=1.06$ nm.  These values are consistent with those previously reported: $D(E_F)=1.5\times10^{18}$ eV$^{-1}$ cm$^{-3}$ and $a=1$ nm.\cite{Knotek}  The transition to variable-range hopping in a strong electric field occurs when the decrease in potential energy $eER(T)$ over a typical hopping distance $R(T)$ becomes comparable to the energy band within which an electron can hop $\Delta\epsilon(T)$ so that phonons are emitted as the electron hops along the field.\cite{Shklovskii}  The critical field $E_C$ for a transition into this regime is given by $eE_CR(T)\approx\Delta\epsilon(T)$, or $E_C\approx k_BT/ea$.  From the curve in \fig{field}, we find the onset of the strong-field regime at $T=58$ K at a critical field of $E_C\approx6.1\times10^6$ V/m from which we find $a\approx0.8$ nm, comparable to $1.06$ nm found above.

In summary, we have demonstrated a contactless measurement of electrical conductance in a thin film of a-Ge.  While we have performed our measurements on a-Ge, this technique is applicable to any thin solid film for which it is impossible to measure current because the contacts are blocking or the film is highly resistive. 

We are grateful to N. Ray for experimental help. This work was supported by the U.S. Army Research Office (911 NF-07-D-0004) and by the Department of Energy (DE-FG02-08ER46515).


\end{document}